# Measurement of segmental mobility during constant strain rate deformation of a poly(methyl methacrylate) glass

Benjamin Bending, Kelly Christison, Josh Ricci, and M.D. Ediger*

Department of Chemistry, University of Wisconsin-Madison, Madison, WI 53706, USA

*To whom correspondences should be addressed

**Abstract :** We describe an apparatus for performing constant strain rate deformations of polymer glasses while simultaneously measuring the segmental mobility with an optical probe reorientation method. Poly(methyl methacrylate) glasses were deformed at $T_g$ - 19 K, for local strain rates between $3.7 \times 10^{-5}$ and $1.2 \times 10^{-4}$ s$^{-1}$. In these experiments, the mobility initially increases in the pre-yield regime, by a factor of 40 to 160, as compared to the undeformed PMMA glass. The mobility then remains constant after yield, even as the stress is decreasing due to strain softening. This is consistent with the view that the sample is being pulled higher on the potential energy landscape in this regime. Higher strain rates lead to higher mobility in the post-yield regime and, for the range of strain rates investigated, mobility and strain rate are linearly correlated. We observe that thermal history has no influence on mobility after yield and that deformation leads to a narrowing of the distribution of segmental relaxation times. These last three observations are consistent with previously reported constant stress experiments on PMMA glasses. The experimental features reported here are compared to computer simulations and theoretical models.

**Introduction**

Polymer glasses are widely utilized materials and their deformation properties often dictate what materials can be used for particular applications, and what material dimensions are required to ensure that mechanical failure does not occur. A fundamental understanding of polymer glass deformation has been difficult to attain for at least three reasons[1–3]. First, molecular rearrangements are required for the glass to deform and, near the glass transition; this molecular motion is incompletely understood even in the absence of flow. Second, a glassy polymer starts in a non-equilibrium state that is changing with time due to structural relaxation (physical aging). Finally, polymer glass deformation is typically well outside of the linear response regime, i.e., the response is not proportional to the amplitude of the mechanical perturbation.

One key idea about polymer glass deformation is that segmental mobility is enhanced during deformation relative to the quiescent glass. Eyring argued that plasticity arises from stress activation of mobility[4]. This concept can be explained in terms of the potential energy landscape that governs the thermodynamics and dynamics of the polymer glass. The application of stress to a solid can be envisioned as tilting the potential energy landscape. In this process, some activation barriers will be lowered thus allowing faster molecular rearrangements. Many models of polymer glass deformation utilize stress activation of mobility, although other models posit the control variable to be free volume[5,6], strain[7], configurational entropy[8,9], or configurational internal energy[10]. For most of these models, the change in mobility (or viscosity) that occurs during deformation is a key source of nonlinearity; the change in mobility does not occur for deformation in a linear response regime. From this perspective, it is enhanced mobility that allows polymer glasses to flow at temperatures where the quiescent glass cannot. The deformation of metallic and colloidal glasses has also been modeled extensively. Interestingly, models developed to describe the deformation of metallic and colloidal glasses typically do not utilize the concept of "mobility" and instead describe "shear transformation zones" or "traps" on the energy landscape[11-16]. It remains to be seen whether these two views can be fully reconciled[17].

Experiments are consistent with the idea that polymer glasses exhibit enhanced mobility during deformation. In early work, it was shown that diffusion of a small molecule in a polymer glass occurred much more rapidly during compression than in the quiescent sample[18]. Solid-state NMR experiments demonstrated increased mobility in the amorphous region of nylon 6 during uniaxial extension[19]. More recently, quantitative molecular mobility measurements during deformation became available as a result of an optical method[20-22] that measures the rotational correlation time of an ensemble of dilute probe molecules. During uniaxial constant-stress deformation, mobility enhancements of more than a factor of 1000 were observed; independent experiments established that the probe reorientation was a good reporter of the polymer segmental dynamics[20]. While these experiments clearly established an Eyring-like regime prior to the onset of flow, the post-flow behavior indicated a more complex relationship between stress and mobility. In this regime, the mobility was found to be roughly proportional to the instantaneous strain rate. These optical experiments also reported a narrowing of the distribution of relaxation times that had not been anticipated by theoretical models[20,23]. Other workers later developed a dielectric relaxation method that also indicated enhanced mobility during uniaxial extension[24].

Molecular dynamics computer simulations support the idea that mobility is enhanced during the deformation of polymer glasses. This work has utilized both atomistic models (polyethylene[25] and polystyrene[26]) and more generic polymer representations[27–29]. Both translational displacements and bond vector reorientation occur more rapidly during deformation with enhancements larger than a factor of 100 being reported. Some of these simulations have shown that multiple measures of segmental relaxation are enhanced by the same factor during deformation[28], giving credence to the use of a simple mobility enhancement to characterize the response of the polymer glass to deformation. Simulations also demonstrate that the potential energy increases during deformation, indicating that deformation can pull the system up the potential energy landscape[26,28,29].

Here we present new optical experiments that measure molecular mobility during constant strain rate deformations of lightly-crosslinked poly(methyl methacrylate) glasses at 373 K ($T_g$ – 19 K). Similar to previous work, the reorientation of a molecular probe is used to indicate changes in segmental mobility. In order to implement constant strain rate deformations, we constructed a new deformation apparatus integrated with a confocal optical microscope to allow measurements of probe mobility. We report constant strain rate deformations where the post-yield local strain rate is between $3.7 \times 10^{-5}$ and $1.2 \times 10^{-4}$ s$^{-1}$. A key advantage of constant strain-rate experiments is the high level of control of deformation in the post-yield state. In contrast to stress-controlled deformations, this allows access to the steady flow regime including strain-softening.

We find that during deformation at constant strain rates between $3.7 \times 10^{-5}$ and $1.2 \times 10^{-4}$ s$^{-1}$, molecular mobility in the PMMA glass initially increases in the pre-yield regime, by up to a factor of more than 100 as compared to the undeformed glass. The mobility then remains constant after yield, even though the stress is significantly decreasing due to strain softening. This is consistent with the view that the sample is being pulled higher on the energy landscape in this regime. Higher strain rates lead to higher mobility in the post-yield regime and, for the range of strain rates investigated, these quantities are linearly correlated. We observe that samples with different thermal histories show identical mobility after yield and that deformation leads to a narrowing of the distribution of segmental relaxation times. These last three observations are consistent with previously reported constant stress experiments on PMMA.

**Experimental Methods**

*Sample preparation.* Lightly cross-linked films of poly(methyl methacrylate) (PMMA) were prepared by thermally-initiated radical polymerization. Methyl methacrylate (Polysciences, Inc.) and the crosslinking agent, ethylene glycol dimethacrylate (Polysciences, Inc.), were purified by removing the inhibitor with an alumina column. 1.5 wt% ethylene glycol dimethacrylate was added to the methyl methacrylate, in addition to the initiator, 0.1 wt% benzoyl peroxide (Polysciences, Inc), and the optical probe N, N′-Dipentyl-3,4,9,10-perylenedicarboximide (DPPC) at a concentration of ~$10^{-6}$ M. The mixture was partially polymerized at 70 °C for 30 minutes to produce a viscous solution; this eased later handling and reduced the volume contraction in the second stage of polymerization. The partially polymerized sample was placed between two 1 x 3 inch glass slides with aluminum foil spacers and clamped using large binder clips. The clamped sample was then further polymerized in an oven under nitrogen, first at 70 ºC for 24 hrs and then at 120 ºC for an additional 24 hrs. These films were removed from the glass slides by ultra-sonication. The glass transition temperature $T_g$ for the PMMA prepared in this manner was 392 ± 1 K; $T_g$ was obtained from the onset value from the 2$^{nd}$ DSC heating scan at a rate of 10 K/min.

The PMMA glass films were cut using a custom steel-ruled die to make 'dog-bone' style tensile samples. This method was based upon the ASTM method D1780-10[30] with slight modifications. Prior to use, each die was inspected under a 20x microscope for any defects on the knife edge. We found that slight defects (caused from mishandling or extended use) cause defects in the sample that later initiate mechanical failure. The die was pressed against the sample by using a Carver hydraulic press or by hand on a PVC self-healing cutting mat. The dimensions of our die are 50% of the ASTM standard to accommodate our smaller deformation apparatus. As shown in Figure 1a, the undeformed film dimensions in the active (center) section are 33 mm long and 2 mm wide. The clip ends of the dog-bone sample are 8 mm wide and 5 mm long with a 5 mm radius of curvature at the shoulder. The clamping procedure for polymerization resulted in films thinner in the middle (around 30 microns) than near the edges (40-50 microns). Yielding generally began in the thinnest part of the film and we found it desirable to have this occur at the film center. By measuring the actual thickness using a micrometer, the cross-sectional area can be determined for use in the calculation of the engineering stress.

*Optical mobility measurement.* To measure the segmental dynamics of PMMA glasses during deformation, we use an optical photobleaching method that has been previously described[20,22]. The reorientation of an ensemble of DPPC molecules is measured using a confocal microscope.

In the undeformed film, the reorientation of DPPC has the same temperature dependence as the segmental dynamics of the PMMA melt as measured by dielectric relaxation[31]. We assume that DPPC continues to be a good reporter of the segmental dynamics below $T_g$ during deformation. To measure probe reorientation we use a linearly polarized laser beam (532 nm) to create an oriented set of unbleached probes. The polarized fluorescence from these molecules excited by a weak circularly polarized beam (also 532 nm) is separated into parallel and perpendicular components. From these two fluorescence intensities we can construct the anisotropy decay function r(t). We fit the anisotropy decays with the stretched exponential function (KWW function):

$$r(t)/r(0) = e^{-\left(\frac{t}{\tau}\right)^\beta}$$

We integrate the correlation function to get the rotational correlation time $\tau_c$. For experiments in which the anisotropy decays by more than 50%, the error in log $\tau_c$ is typically 0.1 and the error in β is 0.1 For experiments in which the anisotropy decays by less than 50%, the β parameter is fixed at 0.31 (the undeformed value) and the error in log $\tau_c$ is typically 0.2.

*Constant strain rate deformation apparatus.* We designed a new apparatus to perform more flexible deformation experiments; our initial instrument could only perform constant engineering stress deformations because it utilized 'dead weight' loading. As shown in Figure 1b, we now use a screw-type linear actuator in order to provide the movement and force for the deformation. This high resolution actuator (Physik Instrumente Gmbh & Co., model M-227-50) has a resolution better than 1 micron and about 40 N of push-pull force. It has a travel distance of 50 mm allowing us to achieve global strains of 1.5 with a maximum strain rate of ~$10^{-2}$ s$^{-1}$. Due to limitations in the optical mobility portion of the experiment, we currently utilize strain rates of

~$10^{-4}$ s$^{-1}$ or less. The deformation apparatus must sit atop a piezoelectric-controlled stage which limits both its mass and the off-axis load. For this reason, we designed the apparatus with a 'U-shape' configuration. In Figure 1b, the direction of the movement is indicated by the red arrows. In order to transfer the movement from the actuator to the sample, the actuator pushes a rigid connection to a single axis cross-roller bearing linear motion stage (THK Instruments); this ensures that the direction of the force applied to the sample remains constant. Attached to the linear motion stage is a miniature load cell (10 N range, XFTC300, Measurement Specialties Inc.). This load cell is a Wheatstone bridge type that has high overload capacity and a high stiffness (greater than $3 \times 10^5$ N/m); the stiffness is important for establishing low instrument compliance. The noise in the amplified signal from the load cell results in ~ 0.02 MPa variation in the measured stress. We periodically calibrate the load cell using standard weights and the absolute force is known better than 1%. We calculate the stress based upon the smallest cross-sectional area of the sample (see Figure 1a). The overall accuracy of our stress measurement is limited by our ability to measure the sample thickness. The error is largest for the thinnest samples (20 microns) and is about 3% in the reported stress.

The sample is connected to a poly(etherimide) rod that is rigidly affixed to the load cell. This allows the load cell to remain at room temperature while the sample is completely surrounded by a temperature-controlled cell that has been described previously[22]. Because our apparatus deforms the sample horizontally, the load cell is subjected to a small amount of off-axis load. Control experiments indicate that this off-axis load has a negligible impact on the accuracy of the stress determination. Since the sample is horizontal and not completely rigid, there is a slight amount of sagging in the sample before the sample is pulled tight due to the deformation. We are able to detect the 'uptick' in the force when the sample is pulled taut; we use this point to establish zeroes for strain and stress. A small window at the bottom of the temperature cell provides access to the optical microscope, allowing measurements of probe reorientation and local strain. The temperature is controlled within 0.2 K and the absolute temperature of the sample is known to about 1 K.

The most difficult aspect of strain-controlled deformation for very thin polymer glass films is proper gripping of the sample. We found it important to apply an even pressure across the entire gripping surface; this prevents portions of the sample under the grips from experiencing high stresses that can cause defects and failure. We previously utilized rubber grips but this results in an instrument compliance that is too high for strain-controlled deformations. For the present experiments, we used 100 grit sandpaper with the paper backing epoxied to brass supports. This provided a low instrument compliance and eliminated almost all slip at the grips, although in a typical experiment a single slip in the range of 5-10 microns is not uncommon.

For a typical constant strain rate deformation, we control the global strain rate by setting the linear actuator speed. We do not control the local strain rate. We measure the local strain by using bleached lines in a 300 micron area of the sample as described in previous publications[20,22]. The global engineering strain (red line) and the measured local engineering strain (black points) for one experiment can be seen as a function of time in Figure 2. Also shown is a fit of the local strain to a 7$^{th}$ degree polynomial (black line). We utilize this fit to represent the local strain as a continuous function in time and we differentiate this function to obtain the instantaneous local strain rate. As shown in Figure 2, the local strain rate is reasonably constant after yield. For each sample, we report the average local strain rate in the post-yield regime. The average local strain rate and the applied global strain rate typically agree to within a factor of 2 or 3.

*Thermal history prior to deformation.* PMMA glass samples were loaded into the deformation apparatus at room temperature. Samples were then annealed well above $T_g$, at 408 K, for at least 2 hours, to provide an equilibrium starting state. Samples were then cooled at 1 K/min to 373 K. Samples were then held isothermally for either 30 or 90 minutes prior to the initiation of a constant-strain rate deformation. After deformation at constant temperature, the sample was again heated to 408 K for at least 4 hours. In this process, the sample retracted to the original dimensions to within a strain value of 0.002, as determined by local strain measurements. A typical sample could be utilized in several deformation experiments prior to failure (usually due to tearing).

**Results**

*Stress-strain behavior.* As described above and shown in Figure 1, an apparatus was designed to allow constant strain rate deformations to be performed on polymer glasses while the segmental mobility was determined via optical measurements. Figure 3 indicates that the mechanical data acquired with this apparatus is accurate and reproducible. For several constant strain rate experiments on PMMA at 373 K, Figure 3 shows the engineering stress as a function of the global strain. These experiments show the expected qualitative features. At very low strains, the stress increases linearly with the strain. This is followed by an anelastic regime, then yielding, and finally a decrease in the stress (strain softening). As described above and shown in Figure 2, while the local strain rate is not controlled in these experiments, it is reasonably constant in the post-yield regime. The average local strain rates for the deformations shown in Figure 3 range from $3.7 \times 10^{-5}$ to $1.2 \times 10^{-4}$ s$^{-1}$.

We can compare a number of quantitative features of the data in Figure 3 to results in the literature. The modulus was calculated by a linear fit to the engineering strain rate up to a global strain value of 0.001. The moduli calculated for the data shown in Figure 3 are $1700 \pm 100$ MPa. Literature results for the modulus of lightly cross-linked PMMA near $T_g$ are 1500 to 2500 MPa[32]. The yield stress (see inset in figure 3) was found to be between 23.5 and 26.6 MPa which is within the range reported in the literature (20-27 MPa)[33,34]. As expected, the yield stress increases linearly with the log of the strain rate[35–37]. For the range of strain rates used here, the global strain at yield is constant at $0.029 \pm 0.002$. The two highest strain rates shown in Figure 3 are very similar and the corresponding stress data (black and red data points) show excellent reproducibility.

*Optical mobility during deformation.* During a constant strain rate deformation, we perform optical mobility measurements by measuring the anisotropy decay for a dilute molecular probe (DPPC). As discussed above, the rotational correlation time for DPPC in PMMA has the same temperature dependence as the dielectric relaxation time for neat PMMA in the equilibrium state. The rotational correlation time for DPPC thus provides access to the polymer segmental relaxation time during deformation.

Figure 4 shows several anisotropy decay functions obtained for DPPC during a constant strain rate deformation on PMMA at 373 K. For this experiment, the average local strain rate was

$6.1 \times 10^{-5}$ s$^{-1}$ and the local strain values at which the anisotropy was measured are indicated. For reference, the black circles show the anisotropy decay for DPPC in the undeformed PMMA glass at the same temperature. The solid lines represent the fitting of the anisotropy data to the KWW function. Initially, the anisotropy curves decrease at a faster rate as the strain increases during the deformation. At higher strains, the anisotropy curves decay at the same rate. In the lower left of the plot, the β values obtained from the KWW fits are shown. The value of β increases during the deformation and then remains nearly constant at higher strain values.

Figure 5 shows the evolution of the rotational correlation time for DPPC in PMMA during six constant strain rate deformations. All the data sets display a common trend. At the beginning of each deformation experiment, the correlation time decreases (i.e., the mobility increases) from its initial value of $10^{4.7}$ s in the undeformed state. At strains of about 0.04, the mobility becomes nearly constant. At the highest values of the local strain rate [$(1.1 - 1.2) \times 10^{-4}$ s$^{-1}$], the mobility has increased by a factor of 160 relative to the undeformed glass.

**Discussion**

*What controls segmental mobility during deformation?* As described in the introduction, theoretical models have made a variety of assumptions as to what controls segmental mobility during deformation. While we do not have access to all quantities of interest in our experiments (e.g., the potential energy of the glass), with this new data we can describe which mechanical properties correlate with segmental mobility in different deformation regimes.

Earlier work with the optical mobility technique provided some partial answers to the question posed by the heading of this section. Those earlier experiments were performed at constant stress and this allowed a detailed investigation of the pre-yield regime. Prior to yield or flow, a very clear correlation between stress and mobility was observed[21]. In this regime, mobility depends exponentially on stress and in this regard the data matches the prediction of Eyring[4]. As those experiments were performed at constant stress, they could not provide access to a steady flow regime. For stresses larger than the flow stress, the sample immediately began to flow and the instantaneous strain rate could vary by more than a factor of 100 in a single experiment. Nevertheless, in this (unsteady) flow regime, the instantaneous mobility was strongly correlated with the instantaneous strain rate.

The experiments reported here provide information about segmental mobility of a polymer glass in the steady flow regime for the first time. We use our new data to test the relationship between mobility and strain rate in steady flow. Figure 6 shows the product of the DPPC rotational correlation time and the instantaneous local strain rate as a function of the local strain for the six deformation experiments shown in Figure 5. The product plotted here is chosen because it approximately corrects for the small variations in the local strain rate (~ 20% in the postyield regime). As shown in Figure 6, the correlation time and the strain rate have a constant product to an excellent approximation in the post-yield regime (alternately, the mobility is proportional to the strain rate). This result is independent of the thermal history of the sample over the range of strain rates tested. A smaller number of deformation experiments have been performed out to larger strain values and one of these is shown in the inset to Figure 6. Clearly the mobility is very nearly proportional to the strain rate over a very large strain window after yield. Over the range of strains shown in the main figure, the stress is not constant, as shown in Figure 3; we return to this point below.

Figure 7 provides an alternate view of the relationship between the rotational correlation time of DPPC and local strain rate. In this plot using data only from the post-yield regime, the average correlation time is plotted against the average local strain rate. Also plotted in Figure 7 are instantaneous correlation times as a function of instantaneous local strain rates from earlier constant stress experiments[20,21]. Both data sets are consistent with a line of slope -1 in the figure, indicating that the mobility is proportional to strain rate in the postyield regime to within the error in the data. Because the constant global strain rate experiments provide an almost constant local strain rate in the postyield regime, these new experiments allow a much more careful investigation of the relationship between mobility and strain rate.

Combining the insights from the constant stress and constant strain rate experiments, we conclude that at an empirical level, stress controls mobility prior to yield or flow, while strain rate controls mobility in the post-yield regime. We note that preliminary constant strain rate experiments extending to lower strain rates show a somewhat more complex relationship between mobility and strain rate. These results will be reported in a future publication.

*Comparison to theory.* Chen and Schweizer have developed a molecular-level Eyring theory to describe the deformation of polymer glasses[38]. In a recent publication, they extended their theory to include the coupled effects of deformation, aging, and thermal history[17,39]. In order to account for dynamics in nonequilibrium states, the theory connects molecular mobility with the amplitude of density fluctuations. In the absence of deformation, the amplitude of density fluctuations decreases until it reaches equilibrium in a process that describes physical aging. Deformation can increase the amplitude of density fluctuations. In this sense, deformation can reverse the effects of physical aging in analogy to "mechanical rejuvenation"; this describes the observation that deformation beyond yield moves the glass to a state that is independent of previous thermal history.

The model of Chen and Schweizer makes predictions for the segmental relaxation time during deformation and we focus our attention on these aspects of the theory. As shown in Figure 3 of reference 39, their model predicts that the segmental correlation time decreases with strain until yield and then remains constant. In this respect, the theory perfectly captures the qualitative features of Figure 6. In the format of Figure 7, the theory predicts that the segmental relaxation time is approximately proportional to $\dot{\varepsilon}^{-0.86}$. Given the narrow range of strain rates over which the new data extends, this dependence describes the data as well as the $\dot{\varepsilon}^{-1.0}$ dependence shown by the solid line. In fact, the absolute segmental relaxation time predicted for PMMA at $T_g - 20$ K by Chen and Schweizer is in quantitative agreement with the data shown in Figure 7.

In light of the close correspondence between the data presented here and the theory of Chen and Schweizer, we offer an interpretation of our results that is heavily influenced by their theory. We interpret the mobility increase in the pre-yield regime as being primarily due to the Eyring mechanism of "landscape tilting". In the pre-yield regime, the instantaneous stress (and the starting state of the glass) determine the segmental mobility. In the post-yield regime, the mobility is constant for a given strain rate even though the stress is decreasing. In the theory, the amplitude of density fluctuations increases throughout this strain softening regime. More generally, we could interpret our experimental results after yield as indicating that part of the stress is being used to pull the system up the potential energy landscape. Higher in the landscape, the barriers between states are smaller, and lower stress levels are required to maintain a constant strain rate. Consistent with this interpretation, in the calculations of Chen and Schweizer, the stress decreases to a constant level at the same strain as where the amplitude of the density fluctuations stops increasing.

Elements of our results are also consistent with a recent model proposed by Fielding, Larson, and Cates[40,41]. Their theory, which aims to describe strain hardening as well as yield, employs a dumbbell and makes use of elements from the Soft Glassy Rheology model. The resulting equation for the segmental relaxation time indicates a steady state solution for constant strain rate, with an inverse relationship between segmental relaxation time and strain rate. This is consistent with the solid line shown in Figure 7.

A recent stochastic constitutive model developed by Medvedev and Caruthers[42] describes the post-yield response of a glassy polymer by incorporating a distribution of relaxation times that stems from the spatially heterogenous dynamics of real polymer glasses[43]. Using parameters for PMMA, this model predicts the mechanical response to constant strain rate experiments for uniaxial compression and extension. In addition, the model predicts that constant strain rate deformation leads to a decrease in the average segmental relaxation time and to a narrowing the distribution of segmental relaxation times in the pre-yield regime. In this model, the distribution of segmental relaxation times becomes constant in the post-yield regime. These three predictions are consistent with the experimental results presented here.

*Comparison to simulations.* As described in the introduction, the deformation of polymer glasses has been studied recently using both atomistic and coarse-grained models. In comparison to experiment, these simulations offer the advantage of providing access to a wider range of observables. On the other hand, because of time scale limitations, computer simulations produce glasses that are trapped much higher on the potential energy landscape than laboratory glasses. As a result, simulations are likely to miss some of the complex features associated with highly cooperative[44] and spatially heterogeneous dynamics[43,45–47]. For example, while constant stress experiments have shown that polymer segmental dynamics during plastic flow become substantially more spatially homogeneous[20], this feature is much less prominent in computer simulations[48].

Riggleman and de Pablo used a coarse-grained model to perform constant strain rate deformations[49]. The segmental mobility (characterized by bond reorientation) increased in the pre-deformation regime and then became constant after yield, in good agreement with Figures 5 and 6. In their simulations, the strain rate was varied by one order of magnitude; their results indicate that the relaxation time varies with strain rate in a manner that is consistent with the Chen/Schweizer prediction, and also with the experimental data shown in Figure 7. Ref 39 reports that the position of the system on the potential energy landscape begins to increase immediately upon deformation. This is not consistent with the interpretation that we have given our experiments and further work is required on this point.

Chung and Lacks have reported molecular dynamics simulations of deformation using a coarse-grained model of polystyrene[50]. They characterized mobility by the translational motion of atoms perpendicular to the deformation direction. They observed higher mobility at higher strain rates and interpreted their results in terms of a "fold-catastrophe" mechanism. In this view, transitions from one potential energy minima to another occur only when the currently occupied minimum ceases to exist. If this is correct, then the segmental relaxation time should be consistent with $\dot{\varepsilon}^{-1.0}$, as shown by the solid line in Figure 7. This view also predicts that the KWW β parameter should not depend upon strain rate. Earlier constant stress experiments are not consistent with this view[20–22,51]. Constant strain rate experiments would need to be performed over a wider range of strain rates than presented here to test this consequence of the fold-catastrophe view.

Warren and Rottler have also reported simulations of a coarse-grained polymer glass using constant stress, constant strain rate, and step strain deformations[52]. They utilize the characteristic time for atomic hopping to characterize segmental mobility in their simulations. They report that deformation enhances mobility and find that their results from different deformation protocols (and different aging times) can be superposed as a function of strain. Although this type of microscopic analysis cannot be done in our experiments, the conclusion reached from these simulations seems consistent with the $\dot{\varepsilon}^{-1.0}$ behavior for the correlation time shown in Figure 7.

**Summary**

We have reported here the first direct measurements of molecular mobility during the constant strain rate deformation of a polymer glass. For a lightly cross-linked PMMA glasses at 373 K ($T_g$ – 19 K), over the range of local strain rates from $3.7 \times 10^{-5}$ and $1.2 \times 10^{-4}$ s$^{-1}$, we observe an increase in mobility as the sample approaches yield followed by constant mobility in the post-yield regime. The mobility increases by a factor of 40 to 160 relative to its value in the undeformed sample. Mobility in the post-yield regime is found to be independent of previous aging time in the range of 0.5 to 1.5 hours. The postyield mobility is consistent with a linear dependence upon strain rate, although it is also consistent with the slightly sublinear dependence predicted by the theory of Chen and Schweizer[39] and observed in simulations by Riggleman et al.[49]. These simulations[49], the theory[39], and our experiments agree that mobility is constant throughout the strain softening regime. A reasonable interpretation of this result is that some fraction of the stress in this regime is being used to pull the system higher on the potential energy landscape.

The apparatus used here for constant strain rate deformations is quite flexible and can perform a variety of other deformation protocols. Constant strain rate experiments cannot distinguish between mobility due to landscape tilting and that which results from pulling the sample up the energy landscape (rejuvenation). Further experiments will explore reversing strain protocols, since this provides a means of differentiating between these two mechanisms of mobility enhancement. Future experiments will explore a wider range of strain rates than those reported here. Preliminary experiments indicate that the trends reported here may not extend to significantly lower strain rates.

**Acknowledgements.** We thank the National Science Foundation (DMR -1104770) for the support of this research. We thank James Caruthers, Leon Govaert, Dan Lacks, and Ken Schweizer for helpful discussions. We thank Lian Yu and Travis Powell for assistance with DSC measurements.

**Figures:**

**Figure 1.** Sample dimensions and the deformation apparatus. Panel a) shows the sample dimensions. Panel b) shows a schematic representation of the deformation apparatus (top view) with the main parts indicated. The apparatus allows constant strain rate, constant stress, and more complex deformation protocols on thin polymer films. The deformation apparatus sits atop an optical microscope to allow probe reorientation measurements during deformation.

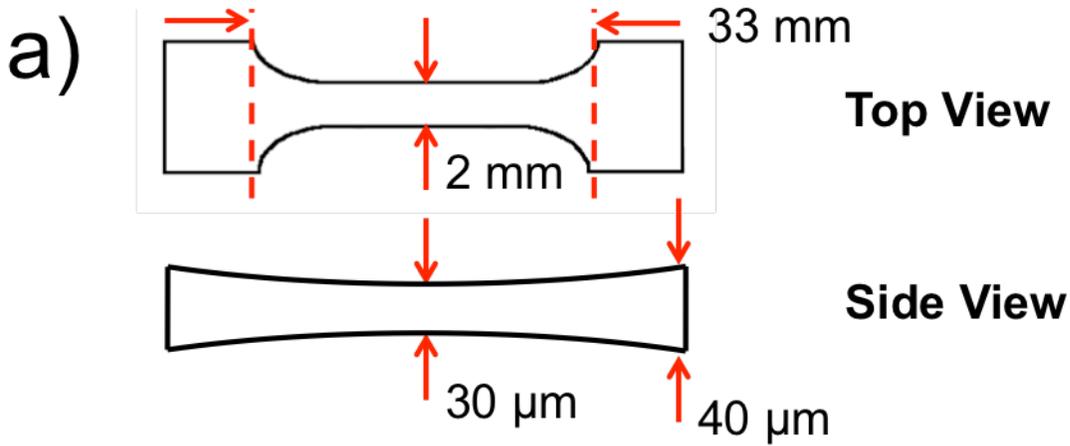

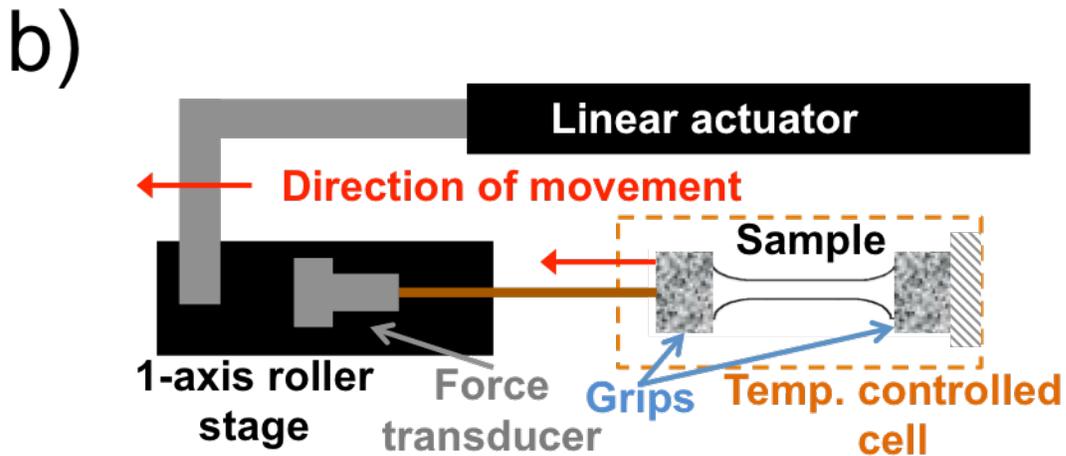

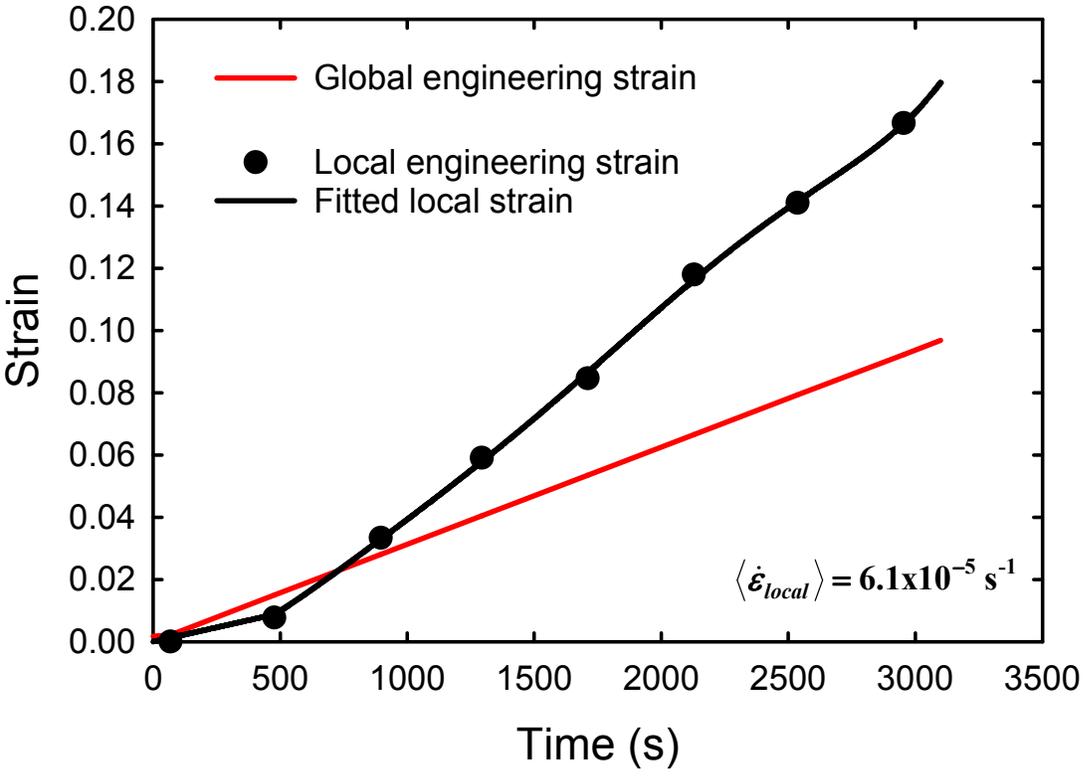

**Figure 2.** Global and local strain for a typical deformation experiment on PMMA glass. The global strain rate is controlled by the linear actuator and is held constant in this experiment. The local strain is measured by the distance between bleached lines. A polynomial fit to the local strain data provides a continuous function of the local strain for further analysis.

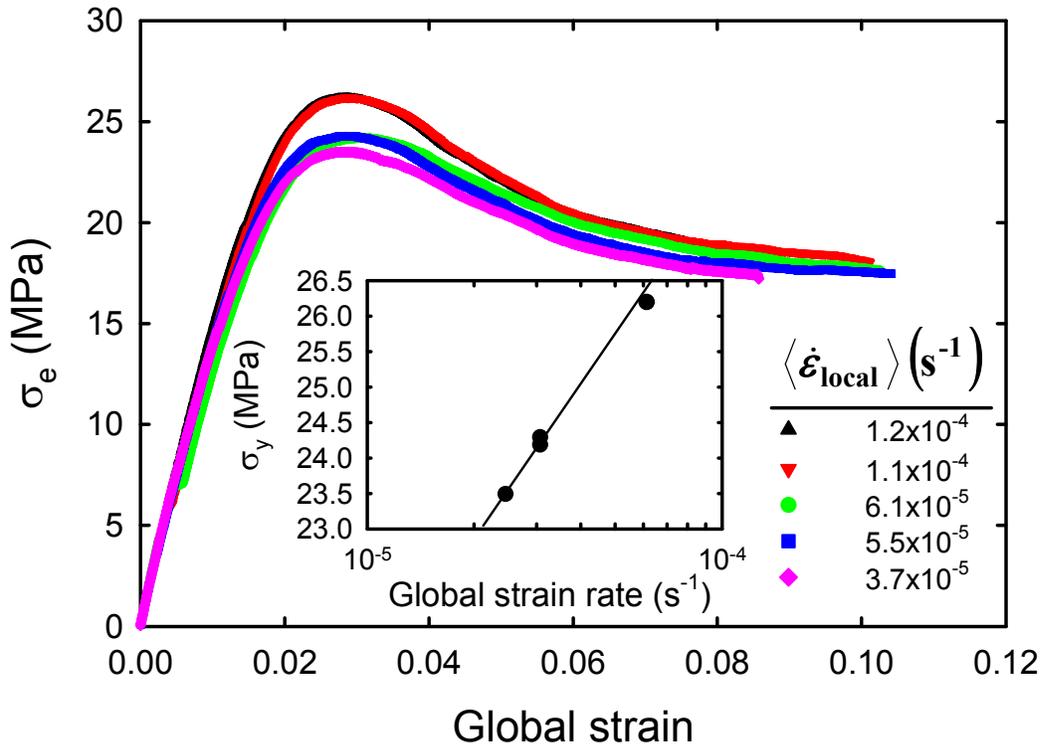

**Figure 3.** Engineering stress as a function of global strain for five constant strain rate deformations on PMMA glasses at 373 K. Each curve is identified by the average local strain rate in the post-yield regime. The inset shows the yield stress as a function of the global strain rate.

**Figure 4.** Representative anisotropy decay functions for the probe molecules (DPPC) measured during a constant strain rate deformation of PMMA at 373 K. For this deformation, the average local strain rate was $6.1 \times 10^{-5}$ s$^{-1}$. The black circles show the anisotropy decay of the undeformed sample. The other four curves show anisotropy decays during the deformation, at the strain value indicated. The anisotropy decays faster during deformation indicating enhanced molecular mobility. The KWW β values are reported for each of the anisotropy decay curves.

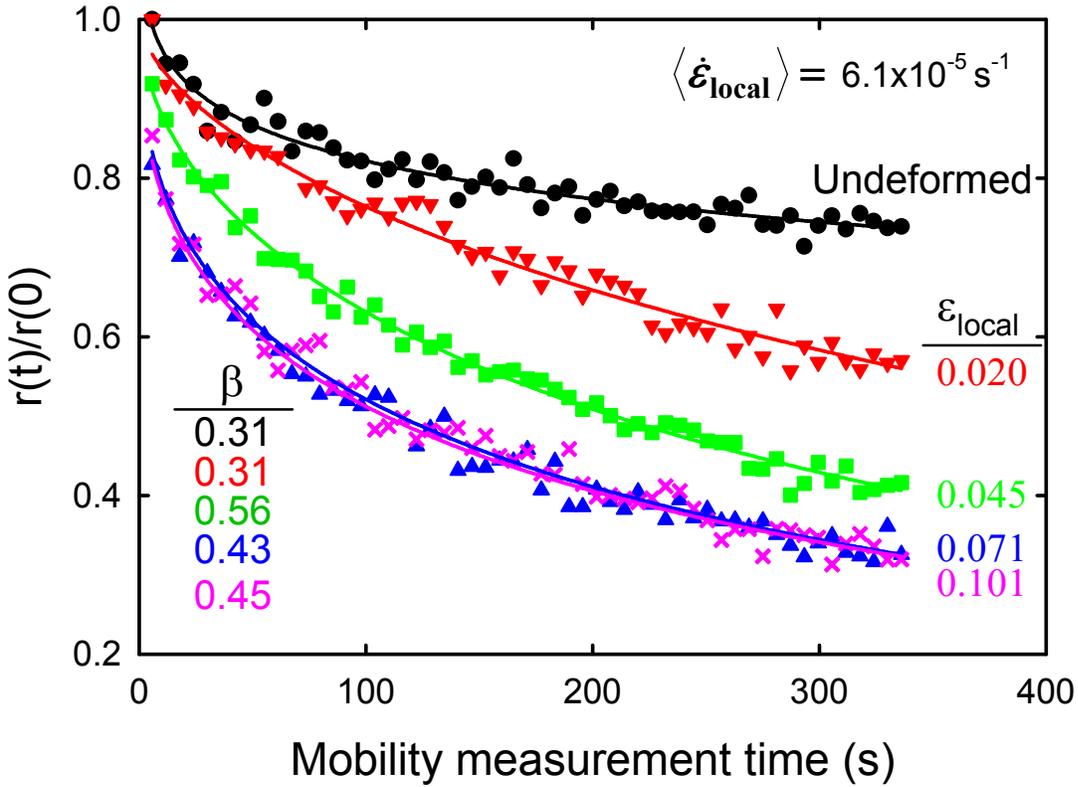

**Figure 5.** The rotational correlation time $\tau_c$ for the molecular probe (DPPC) as a function of local strain during six constant strain rate deformation experiments on PMMA. In each experiment, $\tau_c$ decreases as the deformation begins and then becomes nearly constant after yield. Higher local strain rates result in smaller values of $\tau_c$ and indicate faster segmental mobility. Samples with different thermal histories prior to deformation show similar behavior.

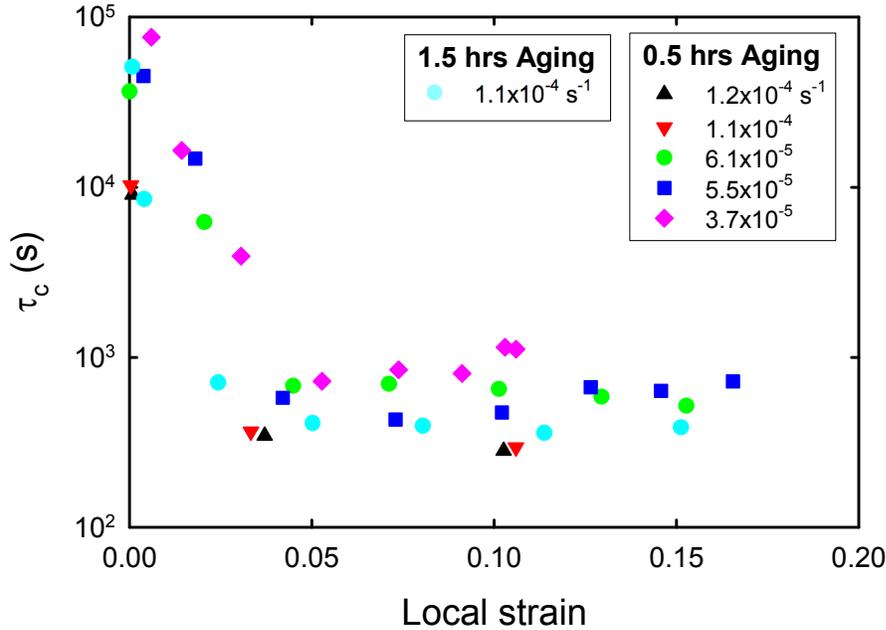

**Figure 6.** Product of the rotational correlation time for DPPC ($\tau_c$) and the local strain rate. In the post-yield regime, this product is very nearly constant as indicated by the vertical dashed line. The inset shows similar behavior for an experiment performed out to higher values of the local strain.

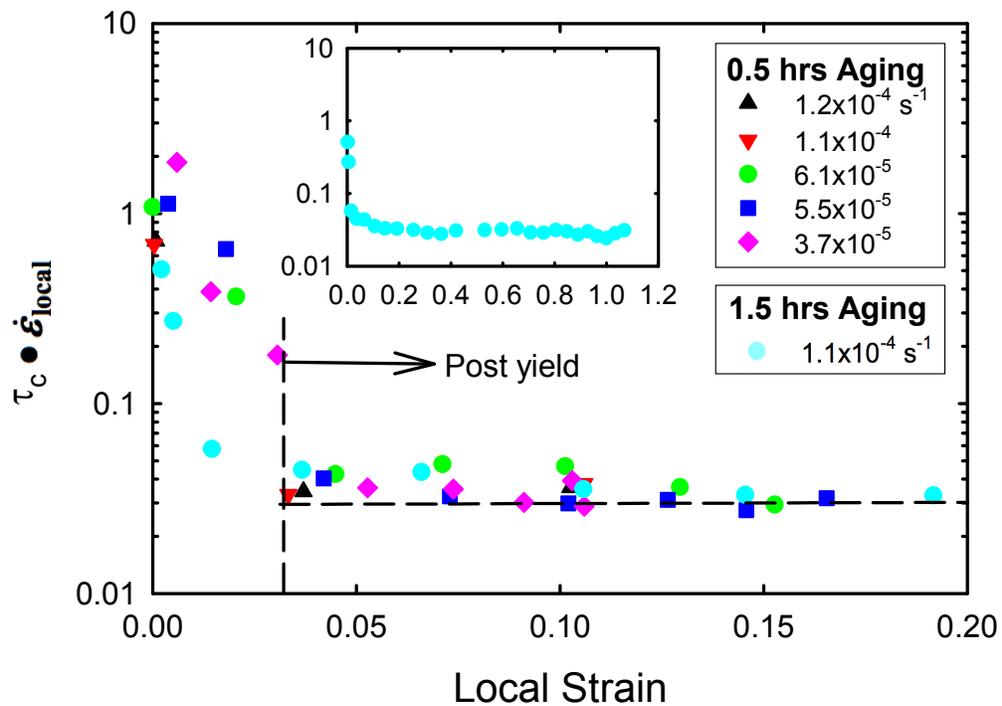

**Figure 7.** Log-log plot of rotational correlation times for DPPC in PMMA glasses as a function of the local strain rate, for the post-yield regime. Results from constant strain rate experiments are shown in blue and red as a function of the average local strain rate. Data in this range of strain rates is consistent with an inverse relationship between correlation time and strain rate as indicated by the solid line with a slope of -1. Error bars represent one standard deviation for the values of the correlation time in the post-yield regime. Shown as open symbols are results reported in ref. 20 from constant stress deformations; for these results, the correlation time is plotted versus the instantaneous local strain rate.

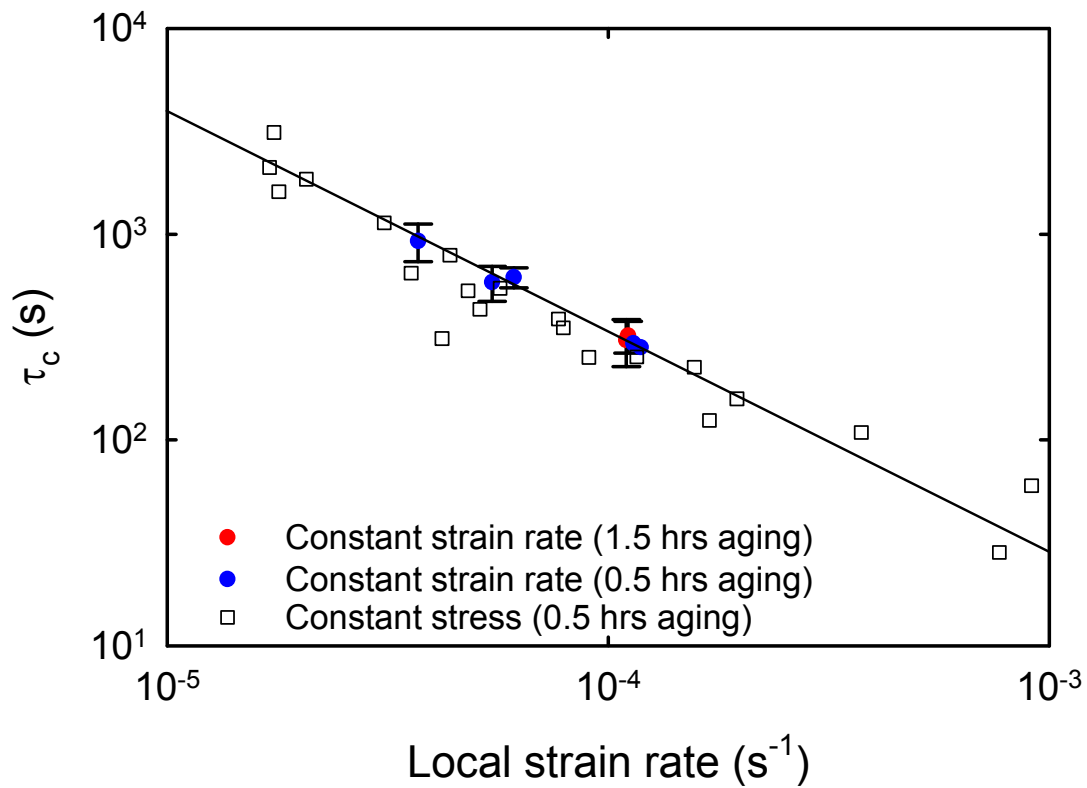

# For Table of Contents Use Only

**Measurement of segmental mobility during constant strain rate deformation of a poly(methyl methacrylate) glass**

Benjamin Bending, Kelly Christison, Josh Ricci, and M.D. Ediger

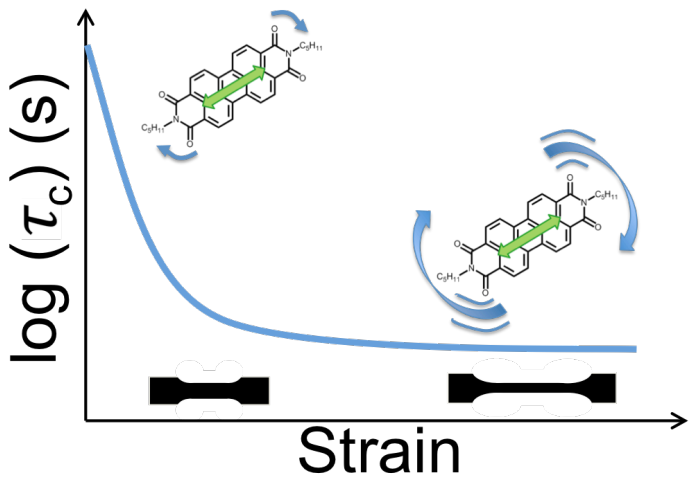